\newcommand{\be}{\begin{equation}}
\newcommand{\ee}{\end{equation}}
\newcommand{\bea}{\begin{eqnarray}}
\newcommand{\eea}{\end{eqnarray}}
\newcommand{\ben}{\begin{eqnarray*}}
\newcommand{\een}{\end{eqnarray*}}
\documentclass[a4paper]{jpconf}
\usepackage{graphicx}
\usepackage{amssymb}
\usepackage{amsmath}
\usepackage{amssymb}
\usepackage{epsfig}
\usepackage{url}
\begin{document}
%
\title{Analysis of the magnetic response of the edge-sharing chain
cuprate Li$_2$CuO$_2$ within TMRG}

\author{Evgeny Plekhanov$^1$, Adolfo Avella$^{1,2}$, and Ferdinando Mancini$^1$}

\address{$^1$ Dipartimento di Fisica ``E.R. Caianiello'' - Unit\`{a}
CNISM di Salerno \\
Universit\`{a} degli Studi di Salerno, I-84084 Fisciano (SA), Italy}
\address{$^2$ Laboratorio Regionale SuperMat CNR-INFM, I-84084 Fisciano (SA), Italy}

\ead{plekhanoff@physics.unisa.it}

\begin{abstract}
It is widely accepted that the low-energy physics in edge-sharing
cuprate materials has one-dimensional (1D) character. The relevant model
to study such systems is believed to be the 1D extended Heisenberg model
with ferromagnetic nearest-neighbor (NN) interaction and antiferromagnetic
next-nearest-neighbor one. Thus far, however, theoretical studies of
such materials have been confined to the case of isotropic interactions.
In the present work, we compare the spin susceptibility of the 1D
extended Heisenberg model with anisotropy in the NN channel, obtained by
means of the Transfer Matrix Renormalization Group method, with that of
the edge-sharing chain cuprate Li$_2$CuO$_2$.
\end{abstract}
\section{Introduction}
In the last decades, an increasing attention has been paid to
low-dimensional materials showing frustrated magnetism. It has been
found, by using
several experimental techniques, that the low-energy physics in such
edge-sharing cuprate materials has one-dimensional (1D) character and
develops in the chains of ${\rm CuO_4}$ plaquettes. A
spin $1/2$ residing on a Cu$^+$ ion interacts with its nearest neighbor
via an oxygen ion in such a way that the Cu-O-Cu bond forms an obtuse
angle and, therefore, the nearest-neighbor (NN) exchange
interaction among Cu$^{+}$ ions appears to be small and ferromagnetic.
The interaction between next-nearest neighboring Cu$^{+}$ ions
develops on a Cu-O-O-Cu path via the exchange mechanism and is
antiferromagnetic. On the basis of these considerations, it is widely
accepted that the effective model for edge-sharing cuprate materials
is a 1D extended Heisenberg model (EHM) with ferromagnetic
NN interaction and antiferromagnetic next-nearest-neighbor (NNN)
one:
\be
   H = -J \sum_i \mathbf{S}_i \mathbf{S}_{i+1}
   + J^{\prime}\sum_i \mathbf{S}_i \mathbf{S}_{i+2}.
   \label{ham0}
\ee
In order to estimate the parameter values for the effective EHM model,
electronic-structure calculations and experimental data from
susceptibility and specific heat are usually combined with theoretical
curves obtained by means of either Exact Diagonalization or
Transfer Matrix Renormalization Group (TMRG).

A comparison between temperature-dependent quantities, such as
susceptibility, specific heat and magnetization, and theoretical
predictions for model~(\ref{ham0}) has already been done for {\it e.g.}
Pb[CuSO$_4$(OH)$_2$] as in Ref.~\cite{kamieniarz_00,baran_00} or for
Rb$_2$Cu$_2$Mo$_3$O$_{12}$ as in Ref.~\cite{lu_00}. Such comparisons
showed a good agreement at high temperatures while, at low temperatures,
the necessity to include inter-chain coupling, which could bring the system
into the 3D regime at $T\to 0$, or additional Hamiltonian terms, like
Dzyaloshinskii-Moriya, has been argued.

Up to now, all attempts to compare theory with experiments have been done
for the {\it isotropic} model~(\ref{ham0}) only. However, the
anisotropy of the interactions in such systems has already been
emphasized as for LiCuVO$_4$~\cite{nidda_00} and for
Li$_2$CuO$_2$~\cite{ohta_00}. From the theoretical point of view,
anisotropy lowers the symmetry of the system and could radically
change the properties of the spin model already at zero
temperature~\cite{Plekhanov_05,Plekhanov_06,Plekhanov_07,Plekhanov_08}.

In the present work, we compare the magnetic susceptibility of
Li$_2$CuO$_2$ compound, taken from
Ref.~\cite{mizuno_00}, with that of the 1D extended anisotropic
Heisenberg model
\be
   H = -J_z \sum_i S^z_i S^z_{i+1} + J_{\bot}
   \sum_i ( S^x_i S^x_{i+1} + S^y_i S^y_{i+1})
   + J^{\prime}\sum_i \mathbf{S}_i \mathbf{S}_{i+2},
   \label{ham}
\ee
calculated within TMRG and try to infer the anisotropy ratio
$J_{\perp}/J_z$ by fitting the experimental curves. For the sake of
simplicity we assume anisotropy only in the NN channel, since in
Li$_2$CuO$_2$ NN coupling is dominant.
\section{Method}
A detailed description of the TMRG method can be found
in~\cite{xiang_00}. In successive diagonalization-decimation of the
transfer-matrix, we have used the Arnoldi algorithm with implicit
restart~\cite{arpack} followed by the bi-orthogonalization of the left
and right eigen-spaces of the non-symmetric density matrix. In doing the
decimation, we retain at most $100$ lowest eigen-states of the
density matrix. We have verified that the truncation error in our
calculations was less than $10^{-4}$. Since the model~(\ref{ham})
contains also the NNN term and the TMRG is designed for systems with
only the NN one, we adopt the mapping of the Hamiltonian~(\ref{ham})
proposed in Ref.~\cite{maeshima}.
\section{Results} 
\begin{figure*}
  \begin{center}
  \begin{tabular}{lr}
  &\begin{minipage}[l]{.45\textwidth}
    \begin{center}
      \epsfig{file=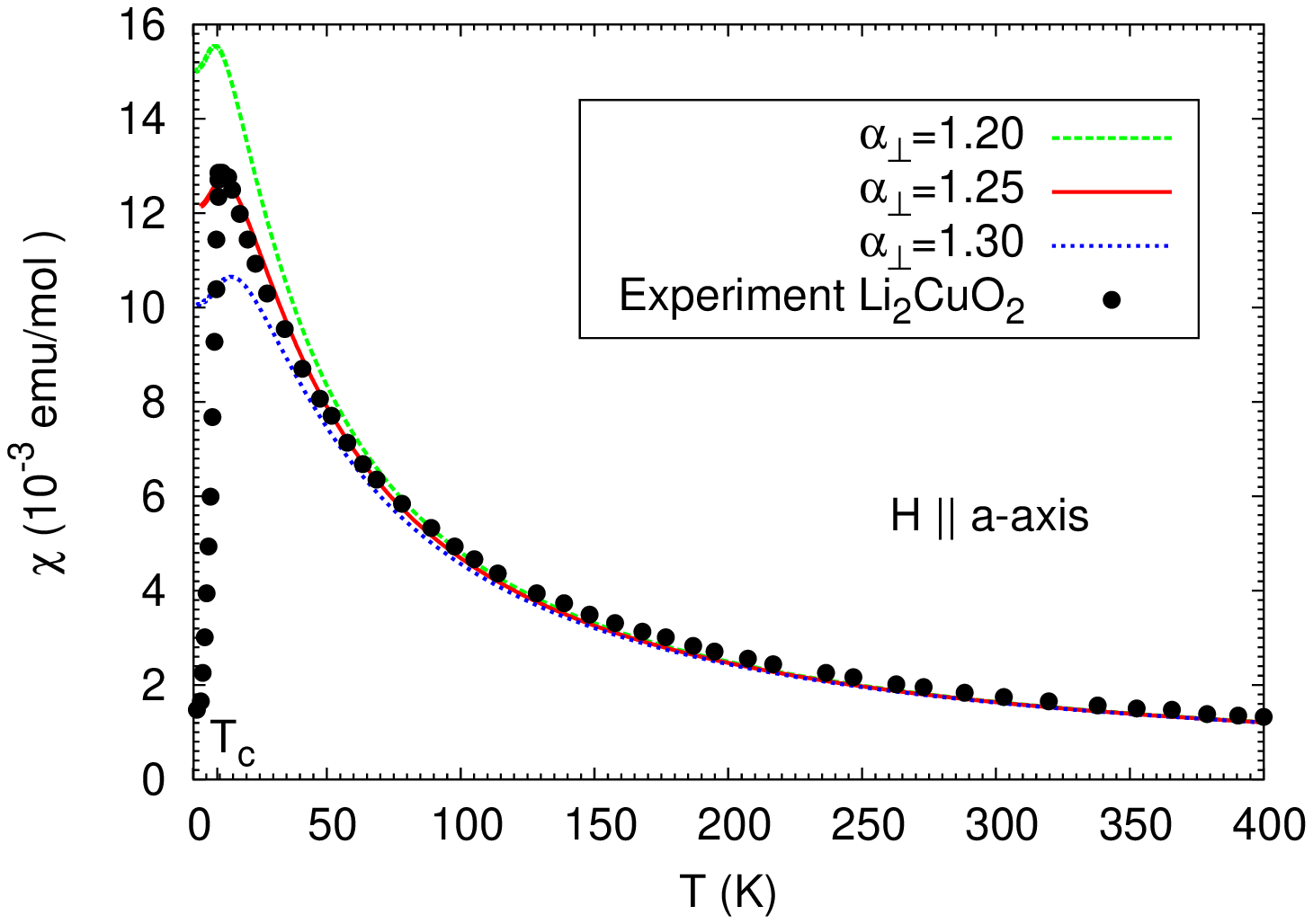, angle=270,scale=0.31}
    \end{center}
  \end{minipage}
  \hspace{0.7cm}
  \begin{minipage}[c]{.45\textwidth}
    \begin{center}
      \epsfig{file=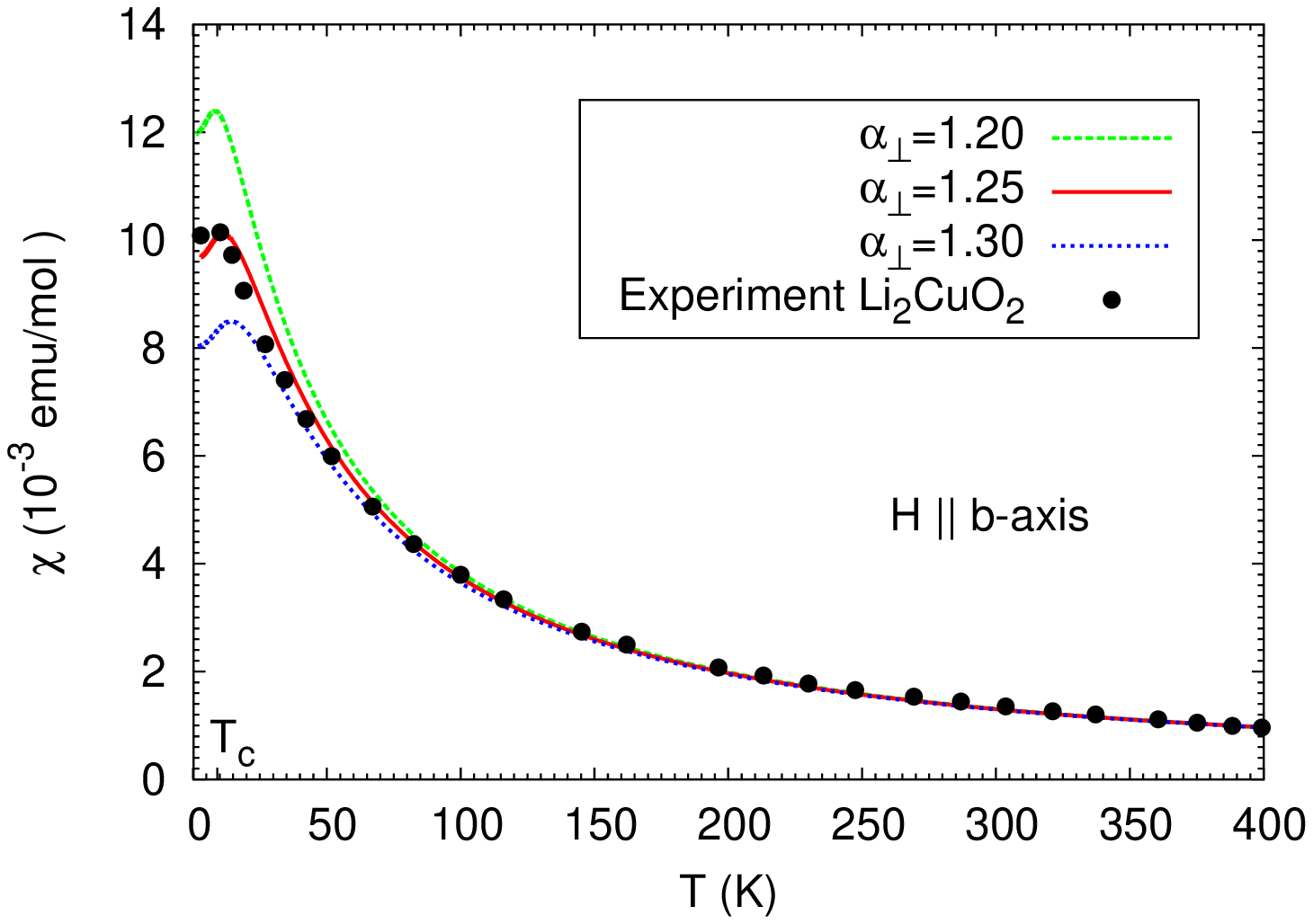, angle=270,scale=0.31}
    \end{center}
  \end{minipage}\\
  &\begin{minipage}[c]{.45\textwidth}
	\begin{center}
	  \epsfig{file=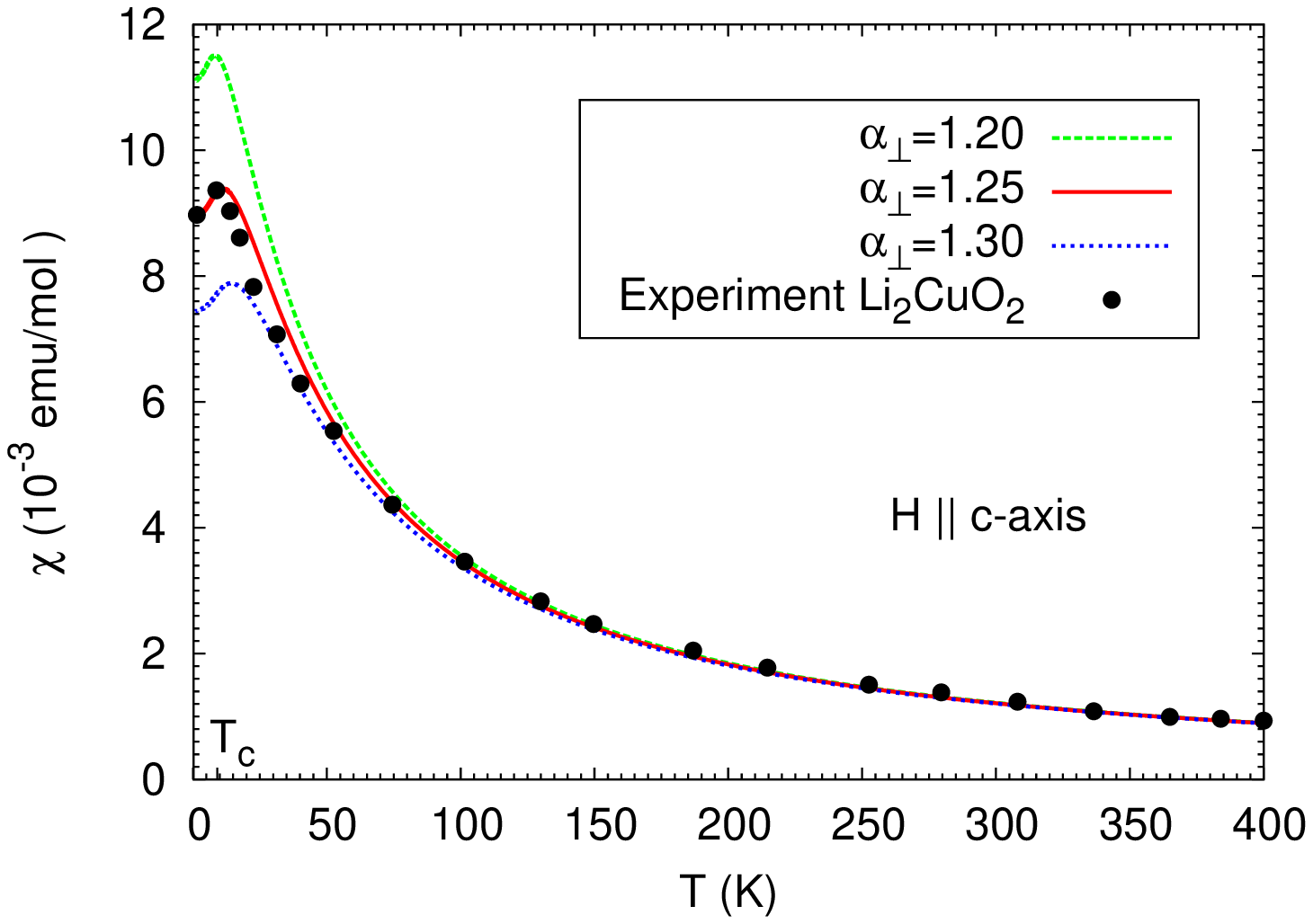, angle=270,scale=0.31}
	\end{center}
  \end{minipage}
  \hspace{0.7cm}
  \begin{minipage}[c]{.45\textwidth}
	\begin{center}
	  \epsfig{file=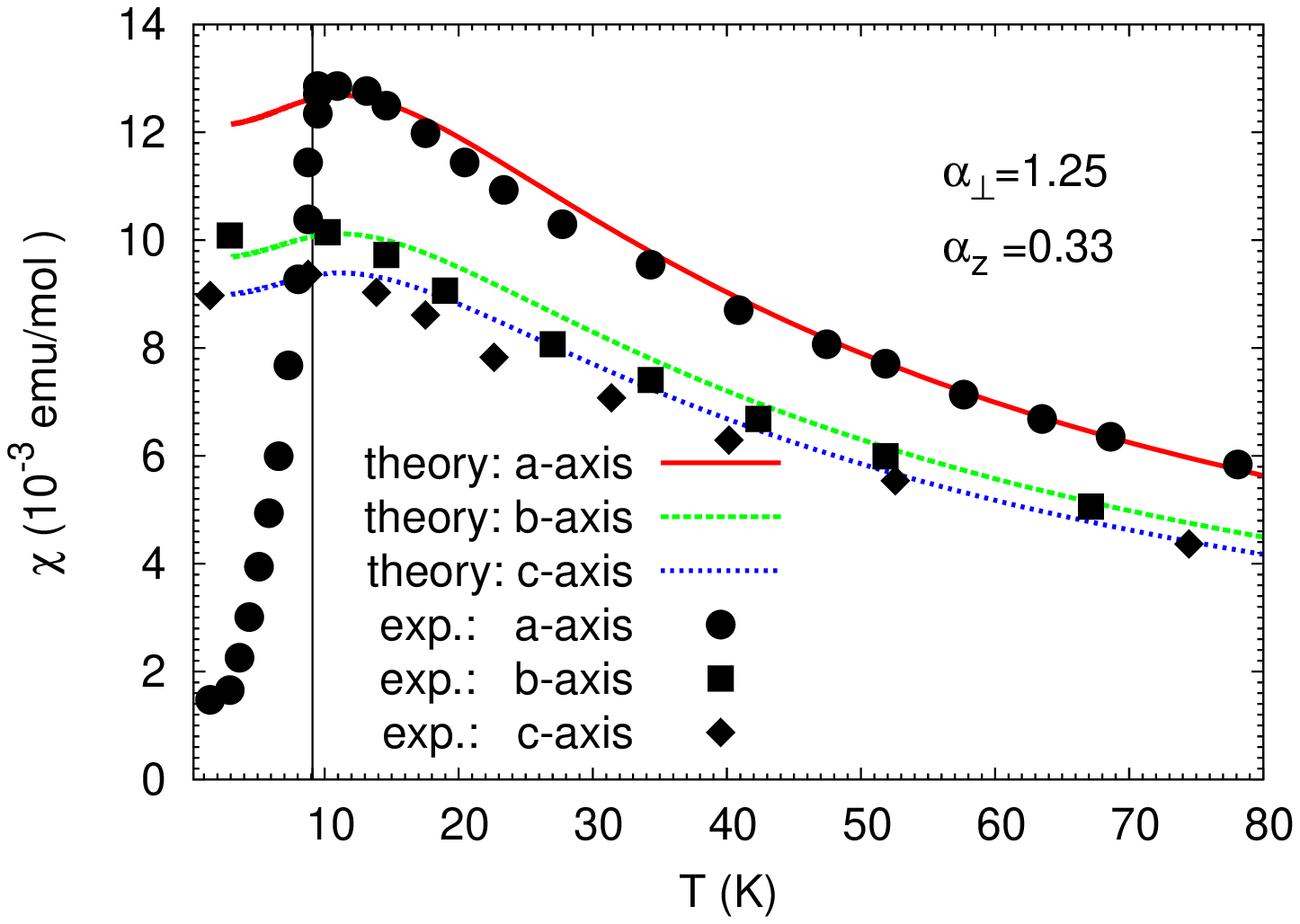, angle=270,scale=0.31}
	\end{center}
  \end{minipage}\\
  \end{tabular}
   \caption{First three panels: theoretical fits for magnetic
   susceptibility of model~(\ref{ham}) to the experimental data for
   Li$_2$CuO$_2$ at $\alpha_z=J_{\perp}/J_z=0.33$, different values of
   $\alpha_{\perp}=J_{\perp}/J_z$ and for different orientations of
   magnetic field. The values of $g$-factors, extracted from Curie's law
   for different field orientations are: $g_a=2.15$, $g_b=1.92$,
   $g_c=1.85$. $J_z=215K$ and $J^{\prime}=0.33J_z$ were taken. Last
   panel: zoomed view in the vicinity of the transition temperature
   $T_c$ of the experimental data and theoretical curves for all three
   orientations of magnetic field.
   }
   \label{fig1}
\end{center}
\end{figure*}

On the basis of the zero-temperature phase diagram obtained for the
model~(\ref{ham}) in Refs.\cite{Plekhanov_07,Plekhanov_08}, we expect
Li$_2$CuO$_2$ to be in a particularly interesting region of the phase diagram
In fact, it could fall either into a massive phase or into a
massless one, depending on the precise values of $J_{\perp}$ and
$J^{\prime}$.
Our strategy is as follows: since the anisotropy estimated from the
experimental data is expected to be small~\cite{nidda_00,ohta_00}, we
first choose $J^{\prime}$ in order to fit best the high-temperature behavior,
then introduce a small deviation of $J_{\perp}$ from $J_z$ and choose
the value of $J_{\perp}$ which fits best the experimental susceptibility
curves at intermediate temperatures. In Ref.~\cite{mizuno_00}, we have
found three sets of data for the susceptibility of Li$_2$CuO$_2$,
corresponding to the three possible orientations of the magnetic field along the
principal axes, as shown in Fig.~\ref{fig1}, being the $b$-axis along
the chain.
 
There is a long and controversial history related to the estimation of coupling
constants $J$ and $\alpha=J^{\prime}/J$ in Li$_2$CuO$_2$, for which
the isotropic model~(\ref{ham0}) is usually assumed and hence
$J=J_z=J_{\perp}$. It was originally believed, after
Ref.~\cite{mizuno_00}, that in Li$_2$CuO$_2$ $J=100K$ and
$\alpha=0.62$. Lately, first from quantum chemistry
calculations~\cite{de_graaf_00} and then from exact diagonalization of
$pd$-Hubbard model, representing finite chains of CuO$_2$
plaquettes~\cite{malek_00}, different estimates came out: $J=142K$,
$\alpha=0.15$ in the first case and $J=146K$, $\alpha=0.23$ in the
second one. Finally, from Density Functional Theory
calculations~\cite{drechsler_03} the values $J=215 K$ and
$\alpha=0.31$ were claimed. We have explored the range of
$\alpha_z\equiv J^{\prime}/J_z\in[0.15,0.62]$ and NN anisotropy
$\alpha_{\perp}\equiv J_{\perp}/J_z\in[0.6,1.3]$. The effect of the
anisotropy on the susceptibility at fixed $\alpha_z$ is to decrease the
height of the intermediate-temperature peak upon increasing
$\alpha_{\perp}$. Since for $\alpha_{\perp}<1$ this peak is
overestimated, (not shown in Fig.1 because its scale is too different) the range
$\alpha_{\perp}<1$ is ruled out and we end up with $\alpha_{\perp}>1$.
We have found that the only set of $\alpha_z$ and $\alpha_{\perp}$
compatible with the experimental data in the widest possible range of
temperatures is the following: $\alpha_z=0.33$ and
$\alpha_{\perp}=1.25$. We have also determined
the values of $g$-factor by requiring that, in the high-temperature
limit, the theoretical susceptibility asymptotically tends to the
corresponding experimental one. This gave us the following values:
$g_a=2.15$, $g_b=1.92$ and $g_c=1.85$.

Below $T_c=9.1 K$, it was found experimentally that in Li$_2$CuO$_2$ 
a long-range order, ferromagnetic along the chain ($b-$axis) and
antiferromagnetic between the chains ($a-$axis) is stabilized~\cite{sapina_00}.
Hence, even a small but finite inter-chain
antiferromagnetic coupling will dramatically reduce the magnetic
response at low temperatures. That is why our theoretical
susceptibility (see Fig.1 last panel) overestimates the magnetic response
of Li$_2$CuO$_2$ at $T<T_c$.
\section{Conclusions} 
In the present manuscript, by using TMRG, we have investigated the
possibility to improve the theoretical description of magnetic
susceptibility in the edge-sharing cuprate material Li$_2$CuO$_2$ upon
introduction of a small anisotropy in the NN channel. The main effect of
the NN anisotropy at fixed $J_z$ appears to be a change in the height of
the susceptibility maximum at intermediate temperatures. Namely, the
increase of anisotropy $\alpha_{\perp}$ suppresses the maximum, while
the decrease of $\alpha_{\perp}$ has the opposite effect. Within the
whole range of possible values of $\alpha_z$, known from the literature,
we found that $\alpha_z=0.33$ together with the anisotropy
$\alpha_{\perp}=1.25$ best describe the experimental susceptibility data
in terms of the model~(\ref{ham}). We believe that taking into account
the inter-chain coupling along $a-$axis would further improve our
description for the corresponding susceptibility at $T<T_c$ and such
work is currently in progress.
\ack
The numerical calculations reported in the present article were done in
part on CINECA CLX cluster (project No. $954$) and on the CASPUR
cluster (project No. $315/09$).
\section*{References}
%

\begin{thebibliography}{10}
\expandafter\ifx\csname url\endcsname\relax
  \def\url#1{{\tt #1}}\fi
\expandafter\ifx\csname urlprefix\endcsname\relax\def\urlprefix{URL }\fi
\providecommand{\eprint}[2][]{\url{#2}}

\bibitem{kamieniarz_00}
Kamieniarz G, Bielinski M, Szukowski G, Szymczak R, Dyeyev S and Renard J~P
  2002 {\em Comp.~Phys.~Comm.\/} {\bf 147} 716 -- 719

\bibitem{baran_00}
Baran M, Jedrzejczak A, Szymczak H, Maltsev V, Kamieniarz G, Szukowski G,
  Loison C, Ormeci A, Drechsler S~L and Rosner H 2006 {\em
  Phys.~Stat.~Sol.~C\/} {\bf 3} 220--224

\bibitem{lu_00}
Lu H~T, Wang Y~J, Qin S and Xiang T 2006 {\em Phys.~Rev.~B\/} {\bf 74} 134425
  (pages~9)

\bibitem{nidda_00}
Krug~von Nidda H~A, Svistov L~E, Eremin M~V, Eremina R~M, Loidl A, Kataev V,
  Validov A, Prokofiev A and A\ss{}mus W 2002 {\em Phys.~Rev.~B\/} {\bf 65}
  13445

\bibitem{ohta_00}
Ohta H, Yamauchi N, Nanba T, Motokawa M, Kawamata S and Okuda K {1993} {\em
  J.~Phys.~Soc.~Jpn.\/} {\bf {62}} {785--792}

\bibitem{Plekhanov_05}
Plekhanov E, Avella A and Mancini F 2008 {\em Acta Phys. Pol. A\/} {\bf 113}
  429

\bibitem{Plekhanov_06}
Plekhanov E, Avella A and Mancini F 2008 {\em J. Optoelectron. Adv. M.\/} {\bf
  10} 1675

\bibitem{Plekhanov_07}
Avella A, Plekhanov E and Mancini F 2008 {\em Europhys.~J.~B\/} {\bf 66} 295

\bibitem{Plekhanov_08}
Plekhanov E, Avella A and Mancini F 2008 {\em {preprint}\/}
  {cond/mat--0811.2973}

\bibitem{mizuno_00}
Mizuno Y, Tohyama T, Maekawa S, Osafune T, Motoyama N, Eisaki H and Uchida S
  1998 {\em Phys.~Rev.~B\/} {\bf 57} 5326--5335

\bibitem{xiang_00}
Xiang T and Wang X {\em Density-Matrix Renormalization: A New Numerical Method
  in Physics\/} ed Peschel I, Wang X, Kaulke M and Hallberg K (New York:
  Springer 1999) p 149

\bibitem{arpack}
   see \verb|http://www.caam.rice.edu/software/ARPACK|

\bibitem{maeshima}
Maeshima N and Okunishi K 2000 {\em Phys. Rev. B\/} {\bf 62} 934--939

\bibitem{de_graaf_00}
de~Graaf C, de~P~R~Moreira I, Illas F, Iglesias O and Labarta A 2002 {\em
  Phys.~Rev.~B\/} {\bf 66} 014448

\bibitem{malek_00}
Malek J, Drechsler S~L, Nitzsche U, Rosner H and Eschrig H 2008 {\em
  Phys.~Rev.~B\/} {\bf 78} 060508

\bibitem{drechsler_03}
Drechsler S~L, Malek J, Nishimoto S, Nitzsche U, Kuzian R, Eschrig H and Rosner
  H 2009 {\em J.~Phys.~Conf.~Ser.\/} {\bf 145} 012060

\bibitem{sapina_00}
Sapina F, Rodriguez-Carvajal J, Sanchis M, Ibianez R, Beltran A and Beltran D
  1990 {\em Sol.~State~Comm.\/} {\bf 74} 779
%
\end{thebibliography}
%
\providecommand{\newblock}{}

\end{document}